\documentclass[times,english,doublespace]{GeotechAuth}

\usepackage{natbib}

\usepackage[T1]{fontenc}
\usepackage[latin9]{inputenc}
\usepackage{amsmath}
\usepackage{amssymb}
\usepackage{graphicx}
\usepackage{esint}
\usepackage{subfigure}
\usepackage{babel}
\usepackage[activate={true,nocompatibility},final,tracking=true,kerning=true,spacing=true,factor=1100,stretch=10,shrink=10]{microtype}
\usepackage{multirow}
\usepackage{booktabs}
\begin{document}
\title{\Large\bfseries
       The critical state of granular media:
       Convergence, stationarity, and disorder}
\author{Matthew R. Kuhn}
%
%
\runningheads{Disorder and the critical state}{M.~R.~Kuhn}
\begin{abstract}
Discrete element simulations are used to 
monitor several micro-scale characteristics within
a granular material,
demonstrating their
convergence during loading toward the critical state,
their stationarity at the critical state,
and the evolution of their disorder
toward the critical state.
Convergence, stationarity, and disorder are studied in
the context of the Shannon entropy and two forms
of Kullback-Leibler relative entropy.
Probability distributions of twenty aspects
of micro-scale configuration,
force, and movement are computed for three topological objects:
particles, voids, and contacts.
The probability distributions of these aspects
are determined at numerous
stages during quasi-static
biaxial compression and unloading.
Not only do stress and density converge to the critical state,
but convergence and stationarity are manifested in all of
the micro-scale aspects.
The statistical
disorder (entropy) of micro-scale movements and strains
generally increases during loading until the
critical state is reached.
When the loading direction is reversed,
order is briefly restored, but continued loading induces
greater disorder in movements and strains
until the critical state is reached again.
\end{abstract}
\maketitle
\section*{Introduction}
The critical state is a foundational concept in geomechanics,
encompassing two 
characteristics of soils and other
granular materials \citep{Schofield:1968a}.
After sustained slow shearing from an initial packing,
a granular material reaches a critical (steady) state in
which the ratio of shear stress to mean stress remains nearly stationary
during further shearing.
Not only does the stress reach a steady condition, but other bulk
measures of internal fabric and structure remain stationary as well
\citep{Thornton:2000a,Pena:2009a}.
Although the values of these characteristics will depend upon
the material's composition
and upon the
form of loading,
for a given material and loading,
a specific steady state condition is eventually attained.
This critical state condition is most enduring at low confining stress,
since
particle breakage at high stress can continue to alter the
internal fabric, even as the shearing stress remains constant
\citep{Luzzani:2002a,MuirWood:2007b}.
Notably, the bulk steady condition
occurs even while
individual particles are undergoing seemingly erratic and highly varied
motions and interactions, notwithstanding a slow, quasi-static loading
\citep{Kuhn:2016b}.
\par
The second characteristic is the convergent quality of
the critical state: the eventual bulk attributes for
a given material are insensitive to the initial particle arrangement
and to prior stress conditions.
For example,
materials that are initially either loose or dense or have different
initial particle arrangements
will approach the same density after sufficient shearing
\citep{Casagrande:1936a,Zhao:2013b}.
These characteristics --- 
persistent convergence toward stationary bulk
attributes ---
resemble those of thermal systems that reliably
approach a stable, equilibrium condition with
sufficient passing of time, even as the underlying
molecular motions remain
dynamic and diverse.
%
\par
In a pair of recent papers, the author proposed a third
characteristic of the critical state:
at the critical state, the micro-scale landscape of
particle arrangements and particle interactions exhibits the greatest
diversity and disorder 
that is consistent with
information known
\emph{a priori} --- information
that constrains the available landscape
\citep{Kuhn:2014d,Kuhn:2016a}.
The two papers, based upon a \emph{conjecture} of maximum
disorder, demonstrated that observed micro-scale features
at the critical state are similar to those that derive from
the conjecture.
These features include distributions of
the contact forces, orientations, and movements
and the topologic arrangements of particles.
\par
Both works deduced this disordered nature
by first considering the full extent (phase space) of potential micro-states
(particle arrangements, movements, forces, etc.) and then
applying available information in the form of reasonable bulk constraints
on the space of micro-states.
Among the many different micro-states that
can satisfy the constraints,
the author sought the specific
collection of micro-states (called a ``macro-state'')
that exhibited the greatest diversity of micro-states.
This approach is equivalent to finding the macro-state
that encompasses the greatest number of micro-states
that satisfy the given constraints.
\citet{Jaynes:1957a} contended that this macro-state
of maximum \emph{entropy}
satisfies the available information but is
otherwise unbiased.
As a simple illustration, consider micro-states of particle 
arrangement, with each micro-state expressed 
as an ordered list of integers, between 1 and 10,
representing, say, the number of contacts of
each particle in an assembly.
Now impose an average of 7.0 for the integers in
a list --- the average coordination number.
Although it is certainly possible that every particle will have seven
contacts, thus satisfying the bulk average, this condition is
highly unlikely and has a low entropy, 
since this macro-state is only realised by
a single list of 7's (a singular micro-state) that
corresponds to a nearly crystalized condition.
Far more lists can be created by permuting equal
occurrences of the numbers 6 and 8,
and the macro-state encompassing such micro-state lists
would indicate a higher entropy and a higher likelihood.
In an analogous manner, the author
applied principles of statistical mechanics (in particular,
the Jaynes formalism of maximising the Shannon entropy)
to derive characterisations of the critical state.
Starting with the assumption of maximum disorder,
the two papers obtained maximally disordered
distributions of various micro-quantities which
compared favourably with distributions
measured in discrete element method (DEM) simulations
at the critical state.
\par
The current work examines connections among the
three characteristics of the critical state: its stationary character,
the convergence tendency, and the tendency toward maximum disorder.
Unlike the author's previous works, which began with an assumption
of maximum disorder at the critical state,
the current work
directly measures disorder and tracks its evolution
in DEM simulations.
The simulations
start from the initial conditions of densely-packed and
loosely-packed assemblies,
continue through
deviatoric
loading in one direction, and finish with
sustained loading in the reversed direction.
With these simulations, the 
disorder in particle movements is shown to
generally increase during loading until
the critical state is reached.
Once attained, all micro-scale measures of fabric, force,
and movement remain in a steady condition.
\section*{Information and disorder}
The intent is to quantify
the diversity exhibited by the micro-states
of a granular material.
For a thermal system,
a micro-state might be presented as
a snapshot of the positions and momenta of the $N$
identifiable constituents (molecules) at a particular instant.
In the current work, snapshots are taken of DEM simulations
at particular strains,
providing the
local densities, coordination numbers,
forces, movements, etc.
A Boltzmann
\emph{micro-state} can be expounded as an ordered $N$-list of such
characteristics
$\kappa$ that characterise
a system's $N$ constituents,
$\{\kappa_{1}, \kappa_{2}, \ldots, \kappa_{N}\}$,
and the full set of all possible lists is the
\emph{phase space} of the system.
Appropriate phase spaces are later identified for depicting
the slow shear-driven movements of particles and contacts.
The \emph{macro-state} to which a micro-state belongs
is represented by the bulk relative frequencies
(empirical probabilities) of items within the micro-state's list.
These frequencies can be extracted for a particular micro-state,
but other micro-states within the phase space
can also share the same frequencies, most
simply as a reshuffling of the order of the $N$ items $\kappa$
within the list.
The mapping of micro-states and macro-states, therefore, is surjective:
each micro-state corresponds to a unique macro-state,
but many micro-states can comprise the same macro-state.
In principle,
constructing a specific micro-state from a more
general macro-state requires
additional information
beyond the gross probabilities, and the amount
of information that must supplement the macro-state,
the Shannon ``missing information,'' is
a measure of a macro-state's diversity or disorder.
The two terms are used interchangeably, and entropy will
designate a quantified measure of disorder.
\par
Methods of quantifying the missing information must accommodate,
in part,
characteristics that are discrete and countable
(quantum states for molecules, or
coordination numbers for a
granular material);
but a general method must also
accommodate information that is
continuous (momenta for classical molecules,
or forces, rates, etc. within a granular material).
With discrete data,
each item $\kappa_{i}$ can, itself, be an $m$-list of discrete
values $n$, $\kappa_{i}=\{n_{1},n_{2},\ldots,n_{m}\}_{i}$, belonging
to the $m$-space of integers $\mathbb{Z}^{m}$.
For example,
the author showed that the topological arrangement of
a two-dimensional assembly of $N$ disks
could be represented as a list (journal) of $N$ integer
pairs, each taken from $\mathbb{Z}^{2}$
\citep{Kuhn:2014d}.
With \emph{continuous} data, the items $\kappa_{i}$ can
be lists of $M$ data points $q$,
$\kappa_{i}=\{q_{1},q_{2},\ldots,q_{M}\}_{i}$,
taken from the continuous $M$-space $\mathbb{R}^{M}$.
We will use both types of data --- discrete and continuous --- to
describe the evolution of disorder during the loading
of granular media.
\par
The disorder 
of a macro-state is measured by first expressing the underlying data
in terms of probabilities $p$ of the component
categories $\kappa$ that appear within micro-state lists.
The Shannon entropy $H$ 
is based upon these
likelihoods --- probabilities $p_{\kappa}$ for discrete systems
and probability density $p(\kappa)$ for continuous systems --- of the
component characteristics $\kappa$ of a macro-state:
\begin{equation} \label{eq:entropy}
\begin{split}
H = -\sum_{\kappa\in\Omega\subset\mathbb{Z}^{m}}p_{\kappa}
     \ln\left(p_{\kappa}\right) \quad\text{or}   \\
H = -\underset{\kappa\in\Omega\subset\mathbb{R}^{m}}{\int}p({\kappa})
     \ln\left(p({\kappa})\right)\,d\Omega
\end{split}
\end{equation}
where phase space $\Omega$ is the set of all possible $\kappa$ components,
$\Omega = \bigcup \kappa$, such that
$\sum p_{\kappa}=1$ and
$\int p({\kappa})\,d\Omega=1$.
%
%
The intrinsic measure $H$ quantifies the diversity
of micro-states
that comprise a macro-state. 
Oddly,
$H$ does not depend upon the values of the underlying
data $\kappa$ but only on their probabilities.
Note the lack of a Boltzmann-like constant in entropy $H$,
conventionally applied
to the Shannon entropy as a scaling factor
to enforce coincidence
with the thermodynamic 
entropy \citep{BenNaim:2008a}.
Two variants of the Shannon measure, described below, will be used
to estimate the
relative disorder
within particular micro-states collected from
DEM snapshots.
\par
Estimating probabilities $p_{\kappa}$ is straightforward
for a micro-state
of a system having a finite number of discrete possibilities
$\kappa$: by simply counting
the occurrences of each condition $\kappa$
and dividing by the length of the data list, $N$.
For the example presented above, entropy $H$ for lists exclusively
constructed from 7's is ${-}1.0\ln (1.0)=0$; for lists of 6's 
and 8's in equal number, $H={-}2(0.50)\ln (0.5)=0.693$.
In general, the more uniform and diverse the distribution, the larger is
its entropy; more biased, uneven, or concentrated
distributions (i.e., those having a more certain outcome)
have lower entropies.
Indeed, the basis of Jaynes' maximum entropy (MaxEnt) principle
is that the most reasonable probability distribution is the one that
is consistent with all available information, but only this
information:  the remaining,``missing'' information is maximised.
For example, when no information is available for the 10 possible
constituents of the previous list, not even an average value, each
constituent should be assumed equally probable,
for which $H=-10\times (0.1)\ln (0.1)=2.30$.
When the average is known to be 7.0, but no other information is
available, entropy $H$ is maximised with a distribution
of exponential form that is biased toward larger numbers and
with the lower entropy of 2.16 (see \citealp{BenNaim:2008a} for other examples).
\par
The entropy measure in Eq.~(\ref{eq:entropy})
is problematic, however, when used to characterise
the disorder exhibited in DEM data.
The Shannon entropy of a continuous system,
as in Eq.~(\ref{eq:entropy}$_{2}$),
suffers from a lack of scale invariance, such that a scaling
of variables $\kappa$ and the corresponding
region $\Omega$ (perhaps with a change in dimensional units)
will alter the entropy $H$ (note the logarithm of density $p$
with its troublesome dimensional units, see \citealp{BenNaim:2008a}).
A second shortcoming is that
$H$ describes the missing information
associated with posterior probabilities $p$ but does not consider
possible access to prior information, even imperfect information,
in the form of preferential probabilities.
\par
These problems are avoided with
an alternative
measure of disorder --- the \emph{relative entropy}
or negative Kullback-Leibler distance
\citep{Melendez:2014a}. 
In this manner, one can account for possible access to
\emph{a priori} information
in the form of the prior probabilities $q_{\kappa}$ or $q(\kappa)$,
which are extracted from past evidence
(such as information gathered at the critical state).
Two sets of prior probabilities are considered.
\par
With the first relative entropy,
it is assumed that no prior information is available
about the particle arrangements, forces, or movements,
so that one can test
whether the disorder of these
quantities increases
as a granular material is loaded.
For this case,
the missing information (or disorder) of the measured posterior
probability $p$ is taken relative
to the most naive, uninformed prior: $q$ is simply the uniform distribution,
$q^{\text{u}}_{\kappa}$ or $q^{\text{u}}(\kappa)$,
spread evenly across phase space $\Omega$.
In this manner,
one measures the disorder of a given system
\emph{relative to the most possibly disordered system},
in which probabilities
are uniformly distributed,
%
\begin{equation} \label{eq:relentropy}
\begin{split}
\mathcal{H}^{\text{u}} = -\sum_{\kappa\in\Omega\subset\mathbb{Z}^{m}}p_{\kappa}
     \ln\left(\frac{p_{\kappa}}{q^{\text{u}}_{\kappa}}\right) \quad\text{or} \\
\mathcal{H}^{\text{u}} = -\underset{\kappa\in\Omega\subset\mathbb{R}^{m}}{\int}p({\kappa})
     \ln\left(\frac{p({\kappa})}{q^{\text{u}}({\kappa})}\right)\,d\Omega
\end{split}
\end{equation}
where $q^{\text{u}}_{\kappa}$ and $q^{\text{u}}(\kappa )$
are proper distributions, each with a net expectation
of~1.
%
The negative Kullback-Leibler distance $\mathcal{H}^{\text{u}}$
is always non-positive.
If a system's density $p$ equals the uniform density $q^{\text{u}}$,
then the
relative entropy $\mathcal{H}^{\text{u}}$ has its largest value of zero,
corresponding to the most disordered system.
Continuing the previous example, if coordination numbers
are between 1 and 10, then $q^{\text{u}}_{\kappa}=1/10$, and the
entropy $\mathcal{H}^{\text{u}}$ for lists exclusively
constructed from 7's is $-1.0\ln (1.0/0.1)=-2.303$; for lists of 6's 
and 8's in equal number, $\mathcal{H}^{\text{u}}=-2(0.50)\ln (0.5/0.1)=-1.609$;
and for a uniform distribution, 
$\mathcal{H}^{\text{u}}=0$.
Again, larger (less negative) values express greater
diversity and disorder.
\par
With the second relative entropy,
one exploits information gained at the critical state, using knowledge
of its
distributions of particle arrangement, force, and movement as
prior information, against which a granular specimen
is compared
as it is being loaded toward the critical state.
This relative entropy simply measures the (negative) distance between
a distribution $p$ and the reference, critical state distribution
$q^{\text{cs}}$,
testing convergence toward (and stationarity of) the critical state:
\begin{equation} \label{eq:relcsentropy}
\begin{split}
\mathcal{H}^{\text{cs}} = -\sum_{\kappa\in\Omega\subset\mathbb{Z}^{m}}p_{\kappa}
     \ln\left(\frac{p_{\kappa}}{q^{\text{cs}}_{\kappa}}\right) \quad\text{or} \\
\mathcal{H}^{\text{cs}} = -\underset{\kappa\in\Omega\subset\mathbb{R}^{m}}{\int}p({\kappa})
     \ln\left(\frac{p({\kappa})}{q^{\text{cs}}({\kappa})}\right)\,d\Omega
\end{split}
\end{equation}
%
This second relative entropy
is a measure of disorder only if one accepts the
critical state as the most disordered system,
a question that is answered
by applying the first relative entropy
$\mathcal{H}^{\text{u}}$ to DEM data.
The scalar $\mathcal{H}^{\text{cs}}$ is similar to the
``state parameter'' of \citet{Been:1985a},
which also characterises
the distance of a current condition from the critical state:
but rather than a difference in bulk void ratios,
$\mathcal{H}^{\text{cs}}$ is the difference between
entire probability distributions of the underlying
micro-scale data.
Although the critical state distributions $q^{\text{cs}}$
could be estimated using
methods of the author's previous papers,
these distributions will be directly extracted from
DEM simulations.
\section*{DEM simulations}
Numerical simulations were used for tracking convergence toward the critical
state and the evolution of disorder
during deviatoric loading.
The simulations were conducted on simple,
two-dimensional assemblies of 676
bi-disperse disks that were contained within periodic boundaries.
Previous work by the author showed that assemblies
with at least a few hundreds of particles
reach nearly the same deviatoric stress
at the critical state \cite{Kuhn:2009b}.
Using two disk varieties, with ratios
of 1.5:1 in size and 1:2.25 in number,
prevented the crystallisation of particle arrangements
that are otherwise observed with mono-disperse assemblies.
To accumulate large data sets, the author created 50 randomly
generated assemblies
(initial micro-states) of dense assemblies and 50 random loose assemblies,
and then slowly
loaded each of the 100 assemblies under identical biaxial
compression conditions.
The creation of these initial assemblies 
is confirmation
that many different configurations, both dense and loose,
are consistent
with a given assembly size and preparation method,
as in the $P$-system of
\citet{Blumenfeld:2009a}.
The particles were isotropically compacted
from a sparse two-dimensional granular gas in which the disks
were assigned random velocities and an artificial
friction coefficient ($\mu=0.20$ for the dense assemblies,
and $\mu=0.50$ for the loose assemblies),
with compaction proceeding until the assemblies ``seized'',
at which the void ratios were
0.175 and 0.246 for the dense and loose
assemblies.
A $0.50$ friction coefficient was then assigned for the subsequent loading
sequence.
\par
\begin{figure}
  \centering
  \includegraphics{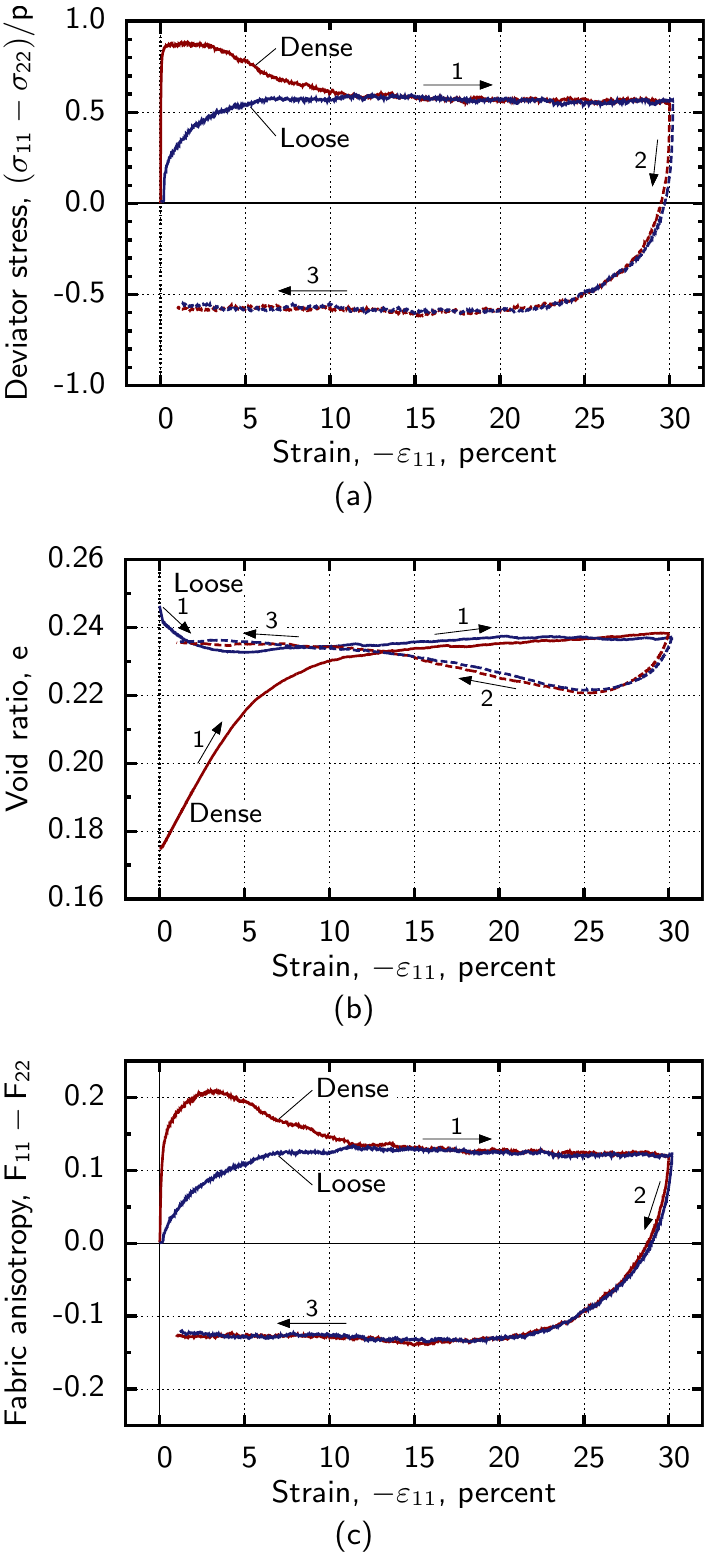}
  \caption{Stress, strain, void ratio, and contact fabric anisotropy
           \citep{Satake:1982a}
           during biaxial loading and unloading,
           averaged from 50 numerical simulations of dense and loose assemblies
           of 676 disks.
           \label{fig:StressStrain}}
\end{figure}

In the first phase of biaxial loading,
the horizontal width of each assembly was reduced (compressed)
at a constant rate
while the height was allowed to expand so that the mean stress within
the assembly remained constant.
Loading proceeded slowly to a horizontal strain of 30\%, 
bringing the
assemblies fully into the critical state of constant stress,
volume, and fabric (Fig.\ref{fig:StressStrain}).
The biaxial loading was then reversed by increasing the width
(i.e. extension) at a constant rate
while reducing the height and maintaining the original mean stress.
Reversed loading was ceased when the strain returned to 1\%.
This reversal phase also brought the assemblies to the critical
state, but with the principal directions of stress, strain, and fabric
rotated by 90$^{\circ}$.
Snapshots were taken of all micro-data at specific strains to
capture the evolving conditions: 22 snapshots during the initial phase
and 21 snapshots during the reversal phase.
By running simulations on 50 assemblies,
the author collected data lists on about $N=29000$ particles,
$N=18000$ voids, and $N=47000$ contacts
for each snapshot of the dense or loose assemblies.
\par
With each snapshot, entropies
$\mathcal{H}^{\text{cs}}$ and
$\mathcal{H}^{\text{u}}$ were computed
for various micro-scale characteristics, described in a
later section,
by using the Voronoi method of \citet{LearnedMiller:2004a},
kernel density estimates of \citet{Duong:2007a},
and direct binning.
%
\subsection*{Bulk convergence, stationarity, and ergodicity}
Figure~\ref{fig:StressStrain} shows changes in the deviator stress and
void ratio during the compressive loading phase and the reversed
loading phase,
as averaged across 50 specimens.
For both dense and loose assemblies, the bulk measures of deviator
stress, void ratio, average fabric anisotropy, and average coordination
number reached a steady, critical state at the strain 20\% during
forward loading (arrow~1, Fig.~\ref{fig:StressStrain}),
and these bulk quantities
also reached the corresponding steady conditions during
reversed loading, after the strain was reduced from 30\% to 10\%
(arrows~2 and~3).
The dense and loose assemblies attained nearly identical deviator
stresses, void ratios, and fabric anisotropies
at the critical state, evidence of the state's convergent
character.
Note that the dense and loose assemblies behaved almost
identically during the reversed loading, with their averaged
stress-strain behaviours being almost indistinguishable.
\par
\begin{figure}
  \centering
  \includegraphics{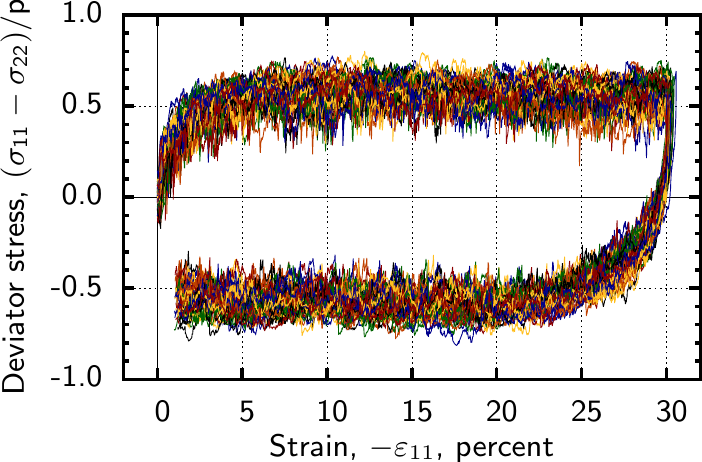}
  \caption{Stress and strain of 50 simulations of loose
           assemblies.
           \label{fig:scatter}}
\end{figure}

Figure~\ref{fig:scatter} shows stress and strain
of all 50 loose specimens,
disclosing the erratic character of stress evolution
(the dense specimens exhibited similar fluctuations).
The fluctuations are also apparent in
averaged data of Fig.~\ref{fig:StressStrain}a as slight,
raspy variations from the average trend.
The 50 specimens were sufficient in number, however,
such that the mean deviator stress $q/p$ has a 95\%
confidence interval of $\pm0.02$.
\par
An analysis of all 50 specimens shows that
after the critical state is reached,
these erratic variations in stress exhibited
both stationarity and ergodicity.
Although both terms are usually applied to time series data,
in the current context,
a monotonically increasing (or decreasing) strain
serves the role of time, and the spatial domain is represented by
the stresses realised in different specimens at the same strain.
Stationarity, meaning constant statistical measures of stress
across a range of strains,
was tested by computing the mean and variance of stress
for the specimens during
two strain intervals: 20--25\% and 25--30\%.
Stationarity was clearly demonstrated at the critical state,
as the means and variances of deviator stress
$(\sigma_{11}-\sigma_{22})/p$ were nearly same
for the two intervals
(means of 0.561 and 0.556,
and variances of 0.0050 and 0.0053).
\par
If a process exhibits the ergodicity property,
multiple instances of the process taken at a given time yield
the same statistical characteristics as that of a single instance
tracked across a sufficiently extended time range.
In the current setting, an ``instance''
is a single simulation and ``time'' is strain.
Ergodicity means that
similar fluctuations would be observed by tracking
a single simulation across a range of strains as would be
observed by noting the results of many simulations at a single strain.
Ergodicity was tested by computing the mean and variance
of all 50 specimens at strain 25\% and comparing these values
to those of individual specimens taken across
a strain range of 20--30\%. 
The statistical measures were almost the same for the two cases,
with nearly identical spatial and temporal means, 
and spatial and temporal variances of 0.0051 and 0.0043.
The statistical characteristics at strain 25\%
are nearly the same as those at
other strains in the range 20--30\%,
which had strength variances of 0.0035--0.0070.
These results indicate that the critical state can be characterised
either by measuring multiple specimens at a given strain or by
measuring a single specimen across a range of strains.
In the current study, multiple specimens were averaged
among multiple strain snapshots at the critical state.
\par
The primary focus of the remainder of this paper is statistical
analyses of micro-scale
(rather than macro-scale, bulk) data,
and as described below,
several tens of thousands of micro-scale samples were used for
this purpose.
It is no surprise that the critical state exhibits
\emph{bulk} stationarity, as the state
is largely defined by this characteristic.
As will be seen, however, stationarity and convergence are more
deep-rooted,
as they pervade all micro-scale aspects as well as being expressed in
bulk, macro-scale characteristics.
%
%
\section*{Micro-scale stationarity, convergence, and disorder}
The evolving micro-scale quantities
are broadly grouped as follows:
%
\renewcommand{\labelenumi}{\Roman{enumi}.}
\begin{enumerate} \setlength\itemsep{0ex}
  \item
    local \emph{configuration} (i.e. packing) quantities,
    such as coordination number, density, and contact orientation,
  \item
    local \emph{force} quantities, such as
    particle stresses and contact forces, and
  \item
    local \emph{movement} quantities, such as 
    particle translations and rotations,
    contact movements,
    and local strains.
\end{enumerate}
\renewcommand{\labelenumi}{\arabic{enumi}}
as presented in the columns of Table~\ref{table:quantities}.
%
The first two aspects are
associated with the status or configuration
of a granular assembly, which are related to the
configuration distributions of
\citet{Edwards:1989a}
or the configurational
entropy of \citet{Valanis:1993a}.
The third set of attributes are transitional rate quantities that are
driven by the bulk deformation and will depend upon the
loading direction.
The distinction between state (being) and transition (becoming) is
explored further below.
\begin{table*}
\centering
\small
\caption{\small Matrix of micro-scale characteristics
         within a granular assembly.
         \label{table:quantities}}
\begin{tabular}{|c|l|p{1.05in}|p{1.05in}|l|}
\cline{3-5}
\multicolumn{2}{l}{} & \multicolumn{3}{|c|}
                         {\rule[-1.0mm]{0mm}{4mm} Local characteristics of a granular assembly}\\
       \cline{3-5}
\multicolumn{2}{l|}{} & I) Configuration & II) Force$^1$ & III) Movement$^2$ \\
\hline
\parbox[t]{3mm}{\multirow{11}{*}{\centering\rotatebox{90}{Object}}}
&Particles & Coordination no.$^3$ & Pressure$^8$ & Horizontal velocity\\
&          &                  & Deviator stress & Vertical velocity\\
&          &                  & Shear stress & Rotational velocity\\
\cline{2-5}
&Voids     & Valence$^4$      &   ---  & Dilation rate$^9$\\
&          & Contact journal$^5$ &        & Compression rate\\
&          & Density$^6$  &        & Transverse shear rate\\
&          &                  &        & Rotation rate\\
\cline{2-5}
&Contacts  & Orientation$^7$  & Normal force     & Slip rate$^{10}$\\
&          &                  & Tangential force & Rolling rate\\
&          &                  &                  & Rigid rotation rate\\
\hline
\multicolumn{5}{|l|}{
\parbox[b]{4.50in}{
\begin{list}{}{%
               \setlength{\itemindent}{0ex}
               \setlength{\labelsep}{0.5ex}
               \setlength{\leftmargin}{1.7ex}
               \setlength{\itemsep}{0.2ex}
               \setlength{\parsep}{0ex}
               \setlength{\topsep}{0ex}
               \setlength{\partopsep}{0ex}
}\small
\item[\textsuperscript{1}]
       forces are normalised by dividing by
       mean stress $p$.
\item[\textsuperscript{2}]
      movements are normalised by dividing by strain increment
       $\Delta\varepsilon_{11}$.
\item[\textsuperscript{3}]
      number of contacts of a particle.
\item[\textsuperscript{4}]
    number of links (contacts) around a polygonal void.
\item[\textsuperscript{5}]
    list of integer pairs
    for reconstructing an entire particle graph
    (see \citealp[\S 2]{Kuhn:2014d}).
\item[\textsuperscript{6}]
    void ratio of a polygonal void,
    ignoring any rattler particles by treating them as void space.
\item[\textsuperscript{7}]
    angular orientation of a contact normal vector.
\item[\textsuperscript{8}]
    particle stresses computed from
    the contact forces (see \citealp{Bagi:1996a}), expressed
    as the pressure $\frac{1}{2}(\sigma_{11}+\sigma_{22})$,
    deviator stress $\sigma_{11}-\sigma_{22}$,
    and shear stress $\sigma_{12}$.
\item[\textsuperscript{9}]
    local deformation rate within a polygonal void,
    computed from the particle movements around
    the void's perimeter (\citealp[\S 2.1]{Kuhn:1999a}).
    The rate is
    a local velocity gradient $L_{ij}$, expressed
    in four combinations: dilation $L_{11}+L_{22}$,
    deviatoric compression $L_{11}-L_{22}$,
    transverse shear $L_{12}+L_{21}$,
    and rotation $L_{12}-L_{21}$.
\item[\textsuperscript{10}]
    relative movements of a particle pair,
    computed from the particles' six component
    rates (two translation rates
    and a rotation rate for each particle).
    These six values are expressed in four combinations:
    three tangential combinations
    (slip, rolling, and rigid rotation, as described in
    \citealp{Kuhn:2004k}) and the relative normal movement
    of the two particles.
\end{list}
}}\\
\hline
\end{tabular}
\end{table*}
\par
The micro-scale of
a two-dimensional granular material can be described
with any of three topological objects:
its non-rattler grains, which can be
represented as the nodes of a particle graph;
the contacts between particles, represented as links between nodes;
and the voids, represented as polygonal loops formed by the nodes
and links \citep{Satake:1992a}.
These objects must be distinguished,
because configuration,
force, and movement are expressed in different
ways among the objects, and the three objects' associated micro-state
lists $\{\kappa_{1}, \kappa_{2}, \ldots, \kappa_{N}\}$
will contain different quantities and
have different lengths $N$.
The three objects are presented in the rows of
Table~\ref{table:quantities}.
\par
Altogether, the table identifies twenty local
micro-scale characteristics that were characterised
in the form of their probability distributions $p_{\kappa}$
at various strain snapshots.
Three of the quantities are discrete
(coordination number, valence, and the contact journal);
the other quantities are continuous.
\subsection*{Convergence and stationarity}
Measure $\mathcal{H}^{\text{cs}}$ in Eq.~(\ref{eq:relcsentropy})
characterises the difference (negative distance) between the probability
distributions of micro-scale quantities during loading and their
eventual distributions at the critical state.
This measure was applied to
each of the twenty micro-scale characteristics in
Table~\ref{table:quantities}.
For the initial loading phase, distributions
were compared with
the reference distribution $q^{\text{cs}}$ at
strain 30\%; during the reversed loading,
the final critical state distributions at strain 1\% were used.
\par
\begin{figure}
  \centering
  \includegraphics{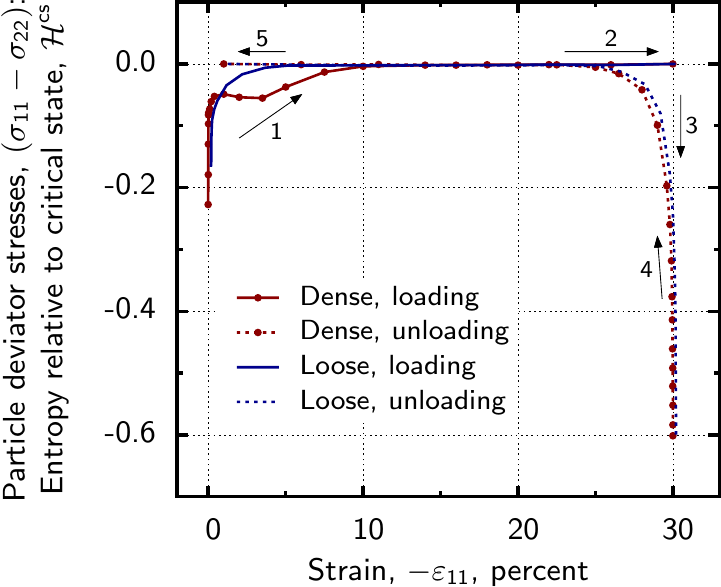}
  \caption{Evolving distance between distributions of the deviator
         stresses within particles and the distribution at the
         critical state.  A value of zero represents perfect
         conformance with the critical state distribution.
         \label{fig:RelCS}}
\end{figure}
\begin{figure}
  \centering
  \includegraphics{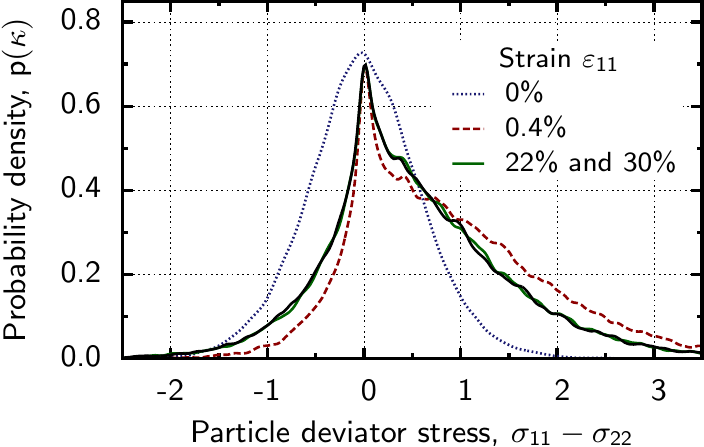}
  \caption{Density distributions of the mean stress within particles
           for the dense specimens during loading at various strains.
           \label{fig:density}}
\end{figure}

As an example of the twenty characteristics, Fig.~\ref{fig:RelCS}
shows the evolution of $\mathcal{H}^{\text{cs}}$ for
the particular
distribution of the local deviator stresses within particles, the difference
$\sigma_{11}-\sigma_{22}$ for each particle within the
$N$=29,000 particle samples.
The bulk stress was initially isotropic,
such that any small deviator stresses (both positive and
negative) within individual particles
balanced to yield a bulk deviator
stress of zero.
This initial
distribution of deviator stress was quite different
than that at the critical state, as is shown in Fig.~\ref{fig:density}:
upon the start of loading, the distribution becomes
broader and more diverse, exhibiting greater disorder.
At the strain 0\%, the (negative) distance $\mathcal{H}^{\text{cs}}$ was
$-0.23$ (Fig.~\ref{fig:RelCS}).
After loading to strain 0.4\%,
the negative distance was $-0.05$,
indicating that the distribution of the particle deviator stresses
had become
closer to that of the critical state.
During further loading
(arrows~1 and~2, Fig.~\ref{fig:RelCS}), the distribution of stress
progressively approached the distribution
of the critical state.
At strains greater than 20\%, the distributions became stationary
and were nearly identical to
the critical state distribution: that is,
$\mathcal{H}^{\text{cs}}$ attained a steady value of zero.
The distributions at both 22\% and 30\%
are plotted in Fig.~\ref{fig:density},
and any small differences can only be discerned
by enlarging the plot.
\par
When the loading was reversed at strain 30\%,
the distribution of deviator stress was very different
from its eventual critical state distribution
(arrow 3 in Fig.~\ref{fig:RelCS}),
since the previous loading
had imparted a positive bias to the particle deviator stresses,
a direction counter to the opposite bias
that was progressively induced during further reversed loading.
The reversed loading quickly brought the distributions
to those of the eventual (reversed) critical state
(arrows 4 and~5 in Fig.~\ref{fig:RelCS}), with
an $\mathcal{H}^{\text{cs}}$ equal to zero.
\par
As with the local deviator stresses,
the other nineteen
characteristics in Table~\ref{table:quantities}
were also considered, and 
the joint probability
distributions of multiple characteristics in each row of the table
were also analysed.
The relative entropies of all micro-scale characteristics
exhibited similar trends:
their distributions converged
to the corresponding distribution of the critical
state, such that $\mathcal{H}^{\text{cs}}$ equalled zero at a (forward)
strain of about 20\% or after a reversal of strain
from 30\% to about 10\%.
The distributions remained stationary during further loading.
That is, convergence and stationarity at the critical state
are manifested in all twenty
micro-scale aspects.
\subsection*{Disorder}
The evolution of micro-scale
disorder was investigated by computing the
relative entropy $\mathcal{H}^{\text{u}}$ of
Eq.~(\ref{eq:relentropy})
for the twenty characteristics in
Table~\ref{table:quantities}.
The author's
original hypothesis was that the disorder of each characteristic
would increase during loading toward the critical state.
Instead,
although $\mathcal{H}^{\text{u}}$
did increase with all local
\emph{movement} characteristics
(column~III of Table~\ref{table:quantities}),
it did \emph{not} necessarily increase with
the configuration or force characteristics in columns~I and~II.
Upon further thought, one recognises that the critical state is not
a fixed condition or status;
instead it is a transitional, active quality,
which has meaning only in the context of a driving deformation.
Consider, for example, the condition of an assembly being
deformed in (forward) biaxial compression
at strain 30\% (see Fig.~\ref{fig:StressStrain}).
One could separately build a \emph{static} arrangement
of disks having the same
configuration and force characteristics as that of the deforming
assembly, but one can ask
whether this static assembly is ``at the critical state''?
Although the question may seem as semantic nuance,
any loading direction other than
a continued forward biaxial loading will divert the assembly from its
current critical
state of deformation.
That is, the critical state is an active condition in which deformation
and movement are essential ingredients.
As will be seen, disorder in the micro-scale
\emph{movements} does increase while
an assembly is approaching the critical state.
On the other hand, notable examples of \emph{reduced} disorder occur among
the configuration and force quantities in
Table~\ref{table:quantities} (columns~I and~II).
For example,
the anisotropy that develops in the contact orientations
during loading (e.g., \citealp{Thornton:2000a}) is a form of increasing order,
since a uniform,
isotropic distribution of orientations exhibits the least order.
Another form of order develops during plastic deformation:
when contact forces reach their frictional limit,
they are constricted by the friction law, which reduces disorder
in the tangential components of the contact forces
(the friction law can
be viewed as a form of constrictive information).
\par
\begin{figure}
  \centering
  \includegraphics{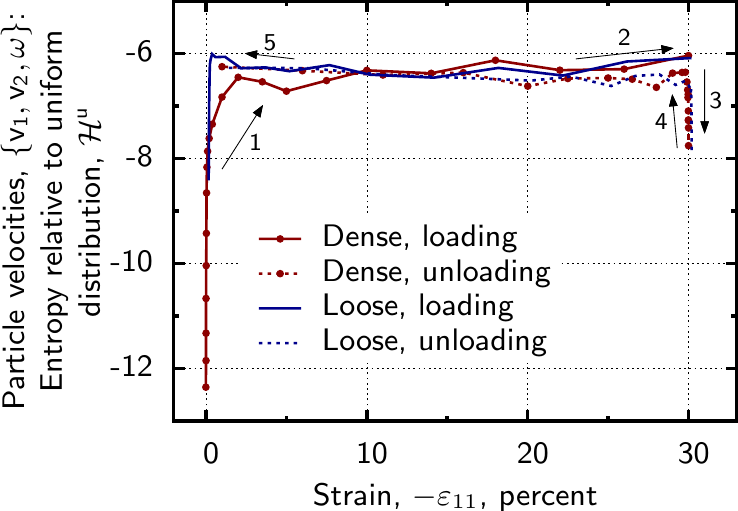}
  \caption{Disorder in the joint probability
           distribution of particle velocities (horizontal, vertical,
           and rotational) as measured with the relative
           entropy $\mathcal{H}^{\text{u}}$.
           \label{fig:entropy_movements}}
\end{figure}

An example of
the tendency toward increasing
disorder in the micro-scale movements is shown
in Fig.~\ref{fig:entropy_movements}.
This plot is of the relative entropy
$\mathcal{H}^{\text{u}}$ of the combined,
joint probability distribution of all three components of particle
movements: horizontal, vertical, and rotational
velocities.
Entropy $\mathcal{H}^{\text{u}}$ is taken relative to
the most disordered (uniform) distribution $q^{\text{u}}$, and a value
of zero represents the greatest disorder.
In Fig.~\ref{fig:entropy_movements}, the movements
of particles within the dense and loose assemblies
are seen to increase from their initially ordered condition
(large negative $\mathcal{H}^{\text{u}}$)
at zero strain
(arrow~1, Fig.~\ref{fig:entropy_movements}) to a larger
value at the critical state (arrow~2).
When the loading is reversed (arrow~3),
the entropy of particle movements is immediately reduced,
corresponding to more ordered (less diverse) movements.
Further reversed loading causes greater disorder (arrow~4) until
the assembly reaches the reverse critical state (arrow~5).
The same trends were measured with all ten movement quantities
in column~III of Table~\ref{table:quantities}: the local,
micro-scale movements
and strains of the particles, voids, and contacts
generally increased as deformation progressed toward the critical state.
\par
\begin{figure}
  \centering
  \includegraphics{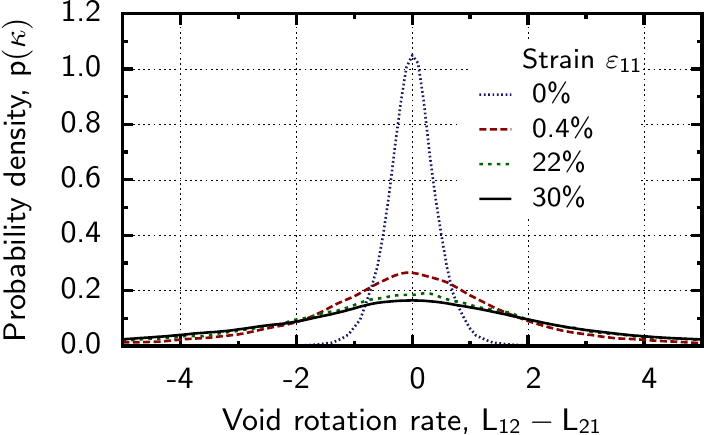}
  \caption{Density distributions of the void rotations within
           the dense specimens during loading at various strains.
           \label{fig:density2}}
\end{figure}

The increasing disorder is also seen in the probability distributions
of Fig.~\ref{fig:density2}.
This figure shows distributions of the rotations (vorticity) of the voids
in the dense assemblies during the loading phase.
A fairly peaked distribution at strain 0\% broadens and becomes
more diverse during loading, attaining a nearly stationary entropy
at the critical state snapshots of 22\% and 30\%.
\par
In some cases,
the assemblies temporarily reached a larger entropy
$\mathcal{H}^{\text{u}}$ prior
to the critical state (for example, during loading at the small
strains of 1\%--3\% in Fig.~\ref{fig:entropy_movements}).
Note that $\mathcal{H}^{\text{u}}$ measures the comparison
of a given probability distribution with the most naive, disordered
system having uniform distribution $q^{\text{u}}$.
Comparisons to the uniform distribution provide a gross
measure of disorder, but $q^{u}$ fails to account for
certain constraints that apply at all stages of loading,
from the start of loading through the critical state.
These constraints include the consistency of particle motions
with the applied bulk deformation, consistency of contact forces
with applied stresses, and energy conservation principles.
Such imposed conditions induce biases in the distributions of
micro-scale quantities that are not considered when comparisons
are made with a uniform distribution.
Previous work has shown that, when considered, these constraints
can account for many of the micro-scale trends that are observed
at the critical state \citep{Kuhn:2016a}.
\section*{Conclusion}
In 1872, Boltzmann published his H-theorem, a rigorous derivation
of an entropy that was based upon micro-scale
analysis of particle collisions
\citep{Boltzmann:2003a}.
Significantly, this probabilistic approach to 
ideal diffuse gases
demonstrated that any temporary disturbance from an equilibrium distribution
is eventually corrected, as the particle assembly naturally returns
to the most probable (i.e. equilibrium) state.
An H-theorem has been proposed for force distributions
within static dense granular assemblies \citep{Metzger:2008a},
but no rigorous proof has yet been presented of a
maximum-entropy principle for dense granular flow.
Although not rigorous confirmation,
the current work and the author's previous works
give support to the concept of maximum entropy
at the critical state.
Simulations of the biaxial loading of simple, two-dimensional
disc assemblies show that micro-scale distributions
of arrangement, force, and movement converge to stationary
distributions.
Furthermore,
disorder in the movements, as characterised by the
entropy relative to a uniform distribution,
tends to increase during loading until the critical state is reached.
That is,
starting from initial packings, the particle motions are
relatively ordered, but as loading
approaches the peak stress and beyond,
the distributions of motions become more
diverse (i.e., disordered).
Upon a reversal of loading --- a temporary disturbance from
the stationary, stable critical state --- the distributions of micro-scale
arrangement, force, and movement abruptly deviate from their
stable condition, but after sustained reversed loading,
the distributions regain the diverse critical state condition.
\par
When one focuses on individual particles or contacts during deformation
of a soil or other granular material,
the motions appear bewildering and complex, with seemingly erratic
fits, stalls, and reversals in all directions \citep{Kuhn:2016b}.
Yet from this local ferment, bulk behavior at the critical state
maintains a steady monotony of stress
and density.
An unsolved and perplexing
problem in granular mechanics is the prediction of
constitutive behavior with models that are based upon the 
complex local interactions of particles.
In the current work, we can glimpse a possible
resolution of the problem, by focusing upon the
probability distributions of micro-scale quantities,
which naturally express their disordered character.
If one accepts the critical state distribution as
the eventual stationary and stable state
and uses, as a generalised ``state parameter'',
the difference between a current distribution
and that of the critical state,
one may be able to predict the progress of
both micro- and macro-quantities during a strain sequence.
Viewing the critical state as a convergent, stationary,
and maximally disorder condition may enable fresh progress
in understanding and predicting soil behavior.
%
%
\section*{Notation}
\begin{tabular}{rl}
 cs & critical state\\
 $e$ & void ratio \\
 $F_{ij}$ & components of fabric tensor\\
 $H$ & Shannon entropy\\
 $\mathcal{H}$ & relative entropy \\
 $\mathcal{H}^{\text{cs}}$ & entropy relative to critical state distribution\\
 $\mathcal{H}^{\text{cs}}$ & entropy relative to uniform distribution \\
 $L_{ij}$ & components of velocity gradient\\
 $N$ & number of particles or objects \\
 $p$ & mean stress \\
 $p_{\kappa}$ & probability of discrete $\kappa$\\
 $p(\kappa)$ & probability density of continuous $\kappa$\\
 $q$ & reference probability distribution\\
 $q_{\kappa}^{\text{cs}}$ & critical state probability distribution, discrete $\kappa$ \\
 $q_{\kappa}^{\text{u}}$ & uniform probability distribution, discrete $\kappa$\\
 $q^{\text{cs}}(\kappa )$ & critical state probability distribution, continuous $\kappa$ \\
 $q^{\text{u}}(\kappa )$ & uniform probability distribution, continuous $\kappa$\\
 u & uniform\\
 $\mathbb{R}^{m}$ & $m$-dimension space, real numbers\\
 $\mathbb{Z}^{m}$ & $m$-dimension space, integers\\
 $\varepsilon_{11}$ & strain in $x_{1}$ direction\\
 $\kappa$ & data values\\
 $\mu$ & friction coefficient at particle contacts\\
 $\sigma_{ij}$ & stress components\\
 $\Omega$ & phase space of possible $\kappa$ values
\end{tabular} 
\bibliographystyle{Geotech}
%
%

\end{document}